\documentclass{article}

\usepackage[preprint,nonatbib]{nips_2018}
\usepackage[utf8]{inputenc} 
\usepackage[T1]{fontenc}    
\usepackage{url}            
\usepackage{booktabs}       
\usepackage{amsmath,amsfonts}       
\usepackage{nicefrac}       
\usepackage{microtype}      
\usepackage{graphicx}
\usepackage{array}
\usepackage{multirow}
\usepackage{adjustbox}
\usepackage{mathptmx}
\usepackage{hyperref}
\usepackage{verbatim}
\usepackage[backend=bibtex,style=alphabetic,maxbibnames=99]{biblatex}
\title{Evaluating Gammatone Frequency Cepstral Coefficients with Neural Networks for Emotion Recognition from Speech}

\author{
  Gabrielle K. Liu\\
  Ravenwood High School\\
  Brentwood, TN 37027 \\
  \texttt{gkml@mit.edu} \\
}

\begin{document}

\maketitle

\begin{abstract}
Current approaches to speech emotion recognition focus on speech features that can capture the emotional content of a speech signal. Mel Frequency Cepstral Coefficients (MFCCs) are one of the most commonly used representations for audio speech recognition and classification. This paper proposes Gammatone Frequency Cepstral Coefficients (GFCCs) as a potentially better representation of speech signals for emotion recognition. The effectiveness of MFCC and GFCC representations are compared and evaluated over emotion and intensity classification tasks with fully connected and recurrent neural network architectures. The results provide evidence that GFCCs outperform MFCCs in speech emotion recognition.
\end{abstract}

\section{\label{sec:1} Introduction}
In recent years, human-computer interactions have become increasingly representative of realistic interpersonal interactions. AI assistants are now able to understand much of the content of human speech. Humans also convey information through emotional cues, and current AI technologies are unable to engage in emotion communication. This gap and the potential benefits arising from emotion artificial intelligence have prompted growing interest in speech emotion recognition. 

Emotion recognition is usually modeled as a classification task. Accordingly, an important question lies in how to best represent a speech signal and capture its emotional content—that is, which features of speech should be used to generate a speech representation? Mel Frequency Cepstral Coefficients (MFCCs) are one of the most commonly used speech features for speech recognition. Existing studies have found that MFCCs lead in performance in comparison to other commonly used speech features (e.g. loudness, formants, linear predictive coefficients) \cite{ref1,ref2,ref3}.

While MFCCs have gained attention in recent years in the context of speech emotion recognition, Gammatone Frequency Cepstral Coefficients (GFCCs) have remained underappreciated. GFCCs are sometimes used in speech and speaker recognition systems \cite{ref4,ref5,ref6}. In contrast to MFCCs which are based upon the Mel Filter Bank, GFCCs are based upon the Gammatone Filter Bank, where the filters model physiological changes in the inner ear and external middle ear \cite{ref8}. Compared to MFCCs, they are more robust against noise are often used in speaker identification systems \cite{ref7,ref9,ref10}.

In this study, we propose that GFCCs are superior to MFCCs for speech emotion recognition. Specifically, we seek to evaluate GFCC representations of speech versus MFCC representations for the tasks of speech emotion and intensity classification. In section~\ref{sec:2}, we describe the experimental setup. In section~\ref{sec:3} we discuss results and conclude by outlining future work.

\section{\label{sec:2} Methodology}
\subsection{\label{subsec:2:1} Dataset}
In this study we analyze the effect of GFCCs versus MFCCs for emotion classification and intensity classification. We use the Ryerson Audio-Visual Database of Emotional Speech and Song (RAVDESS) \cite{ref11}. RAVDESS encompasses eight emotions (anger, calm, disgust, fear, happiness, sadness, surprise, neutral), all of which aside from neutral are expressed at two intensities (normal, strong). There are 12 female speakers and 12 male speakers with 60 speech files each (4 files for each emotion-intensity pair and 4 files for neutral). Speech files have annotated emotion and intensity labels, are less than fifteen seconds long, and consist of a single sentence expressing a single emotional state and read in North American English. All classes are balanced.

\subsection{\label{subsec:2:2} Speech representations}
MFCCs and GFCCs are generated from a given speech signal through a series of transformations (Table~\ref{tab:table1}). The first thirteen MFCCs are known to provide unique characteristics of a signal and closely model changes in vocal tract shape during human speech production \cite{ref2}. The deltas and double deltas of these thirteen coefficients are often used to further capture temporal information.

\begin{table}[ht]
\centering
\begin{tabular}{p{1.6in}p{1.6in}p{1.6in}}
\hline\hline
\textbf{Operation}& \textbf{MFCC} \cite{ref12,ref13} & \textbf{GFCC} \cite{ref9}\\\hline
\textit{Pre-emphasis} & Improve signal-to-noise ratio& Improve signal-to-noise ratio\\\hline
\textit{Framing} & Divide signal into frames & Divide signal into frames\\\hline
\textit{Windowing} & Hamming & Hamming\\\hline
\textit{Discrete Fourier Transform} & Convert from time to frequency domain & Convert from time to frequency domain\\\hline
\textit{Filter Bank} & Mel Filter Bank & Gammatone Filter Bank\\\hline
\textit{Logarithmic Compression} & Align  filter outputs with human-perceived loudness & Align  filter outputs with human-perceived loudness\\\hline
\textit{Discrete Cosine Transform} & Convert back to time domain & Convert back to time domain\\
\hline\hline
\end{tabular}
\caption{\label{tab:table1}MFCC and GFCC generation.}
\end{table}

\begin{figure*}[ht]
\centering
\includegraphics[width=.4\linewidth]{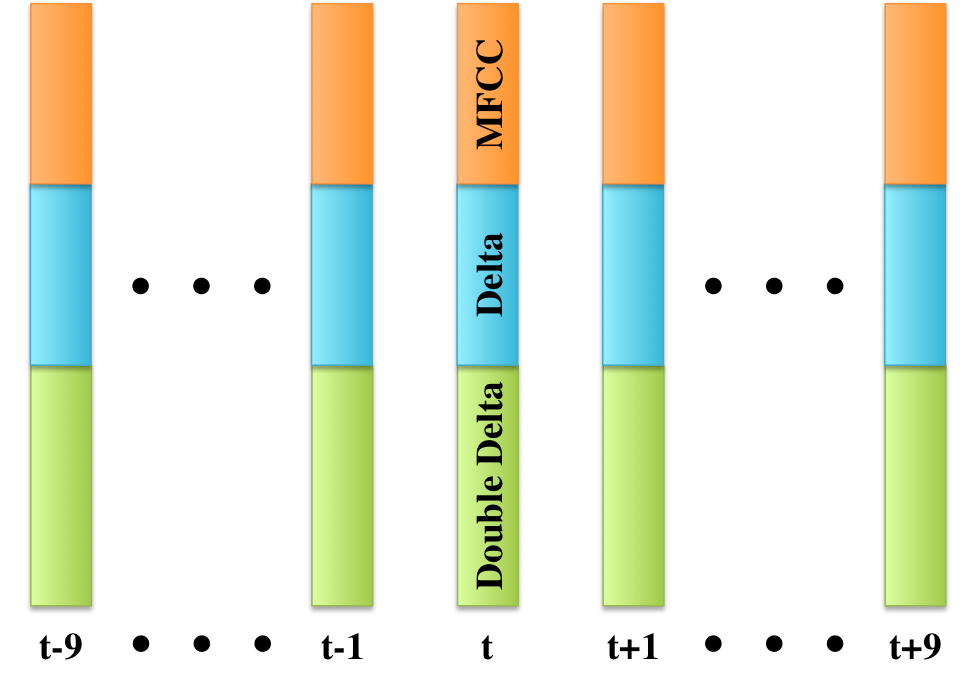}
\caption{\label{fig:fig1}Generation of MFCC representation vector for frame $t$.}
\end{figure*}

For each speech file we generate both a context-based MFCC representation and a context-based GFCC representation on a frame-by-frame basis. For example, to generate the MFCC representation for frame $t$, we compute for each frame $t-9,t-8,\ldots, t-1, t, t+1,\ldots, t+8, t+9$ a vector of 39 MFCCs (13 coefficients with deltas and double deltas). The final vector representation for frame $t$ is the concatenation of these nineteen vectors in sequential order, as shown in Fig.~\ref{fig:fig1}. The GFCC representation for each frame is generated similarly.

\subsection{\label{subsec:2:3} Models}
We compare the effectiveness of MFCC and GFCC speech representations by evaluating the two tasks over various neural network architectures and sizes. We investigate Fully Connected Networks, LSTMs, and Attention-based LSTMs. (Attention is a learned weighting mechanism that allows a neural network to learn to estimate the relative importance of information at different times in a sequence \cite{ref13,ref14}.) Each architecture concludes with a softmax layer for classification to one of eight labels for emotion recognition or one of two labels for intensity classification. Tables~\ref{tab:table2} and~\ref{tab:table3} outlines all experimental architectures in this study.

\subsection{\label{subsec:2:4} Evaluation}
For all experiments, we use batches of 60 samples with the Adam optimizer for batch gradient descent. Rather than standardizing the number of epochs during training, we use early stopping to end training once accuracy does not improve by at least 0.0005 within 15 epochs. Thus, each model is potentially trained for a different number of epochs but yields the best performance. For Fully Connected Networks, we use 20\% dropout and Sigmoid activation for each hidden layer. For LSTM and Attention-based LSTM models, we do not use dropout and use hyperbolic tangent activation, and all speech sequences are padded to 820 frames.

We random sample 75\% of the RAVDESS speech corpus to create our training set, with an equal number of samples per emotion class and intensity class. The remaining 25\% is used to generate the test set, which is also balanced among emotion and intensity classes. All data is normalized by removing the mean and scaling to unit variance. For each architecture, we train models based on MFCC and GFCC representations. In all cases, evaluation is based on a comparison of overall test set accuracy; we define the better model as that which has higher accuracy. 
\section{\label{sec:3} Results and discussion}
Test set results for all experiments are displayed in Tables~\ref{tab:table2} and~\ref{tab:table3}. We observe that the GFCC representation results in higher accuracy over the MFCC representation across all architectures. GFCCs provided an average increase in accuracy over MFCCs of 3.6\% for emotion classification over the architectures we studied. GFCCs consistently outperformed MFCCs in every architecture studied for both emotion and intensity classification, based on equally sized MFCC and GFCC representations. 
\begin{table}[ht]
\centering
\begin{tabular}{c>{\raggedright\arraybackslash}p{1in}l>{\centering\arraybackslash}p{.3in}|>{\centering\arraybackslash}p{.4in}>{\centering\arraybackslash}p{.4in}|>{\centering\arraybackslash}p{.4in}>{\centering\arraybackslash\bfseries}p{.4in}}
\hline\hline

\multicolumn{4}{c|}{} & \multicolumn{2}{c|}{MFCC}& \multicolumn{2}{c}{GFCC}\\\hline
&\multicolumn{1}{c}{Model} & \multicolumn{1}{c}{Activation} & \multicolumn{1}{c|}{Dropout} & \multicolumn{1}{c}{Loss} & \multicolumn{1}{c|}{Accuracy} & \multicolumn{1}{c}{Loss} & \multicolumn{1}{c}{Accuracy}\\\hline

\multirow{6}{*}{\rotatebox[origin=c]{90}{FCNN}\hspace{.1cm}}&F(800)	& Sigmoid&	0.2	&0.763	&0.716&	0.845&	0.731\\
&F(400)	&Sigmoid&	0.2&	0.767&	0.708&	0.791&	0.725\\
&F(200)	&Sigmoid&	0.2&	0.772&	0.699&	0.770&	0.716\\
&F(800)/F(800)&	Sigmoid&	0.2&	0.817&	0.728&	1.019&0.756\\
&F(400)/F(400)	&Sigmoid&	0.2&	0.761&	0.721&	0.837&0.749\\
&F(200)/F(200)	&Sigmoid&	0.2&	0.759&	0.710&	0.747&0.740\\\hline
\multirow{3}{*}{\rotatebox[origin=c]{90}{LSTM}\hspace{.1cm}}&L(800)&	tanh&	0&	1.357&	0.483&	1.375&	0.515\\
&L(400)	&tanh&	0&	1.467&	0.461&	1.373&	0.495\\
&L(200)	&tanh&	0&	1.449&	0.458&	1.479&	0.467\\\hline
\multirow{4}{*}{\rotatebox[origin=c]{90}{LSTM+A}\hspace{.1cm}}&L(800)/A&	tanh&	0&	0.964&	0.689&	1.142&	0.706\\
&L(400)/A&	tanh&	0&	0.864&	0.748&	0.813&	0.768\\
&L(200)/A&	tanh&	0&	0.817&	0.720&	0.686&	0.773\\
&L (100)/A&	tanh&	0&	0.953&	0.715&	0.818&	0.720\\
\hline\hline
\end{tabular}
\caption{\label{tab:table2}Comparison of emotion classification results for GFCC and MFCC representations. Here F(800) denotes a neural network with one 800-unit fully connected hidden layer. F(800)/F(800) denotes a network with two 800-unit fully connected hidden layers. L(800) denotes a network with an 800-unit LSTM layer. L(800)/A denotes a network with an 800-unit LSTM layer with Attention.}
\end{table}

This has several important implications and provides evidence in support of our proposal that GFCCs are a better representation than MFCCs for emotion and intensity classification. An explanation for this result might be found in the differences in how MFCCs and GFCCs are generated for a given audio signal. While both representations model human sound perception to some degree, GFCCs are better aligned to capture the motion of the basilar membrane in the cochlea during hearing \cite{ref8}. We speculate that GFCCs can more closely model the physical changes that occur within the ear during hearing and are therefore more representative of the human auditory system than MFCCs. This notion provides a basic intuition for why GFCCs lead to more accurate emotion and intensity classification than MFCCs. It further suggests that audio signal representations that better align with specific processes in the ear during sound perception can lead to further improvements in speech emotion recognition. This is a potential avenue for future work: to investigate other biologically inspired audio representations and to investigate the effect of creating hybrid representations that draw from multiple such representations.

\begin{table}[ht]
\centering
\begin{tabular}{c>{\raggedright\arraybackslash}p{1in}l>{\centering\arraybackslash}p{.3in}|>{\centering\arraybackslash}p{.4in}>{\centering\arraybackslash}p{.4in}|>{\centering\arraybackslash}p{.4in}>{\centering\arraybackslash\bfseries}p{.4in}}
\hline\hline

\multicolumn{4}{c|}{} & \multicolumn{2}{c|}{MFCC}& \multicolumn{2}{c}{GFCC}\\\hline
&\multicolumn{1}{l}{Model} & \multicolumn{1}{c}{Activation} & \multicolumn{1}{c|}{Dropout} & \multicolumn{1}{c}{Loss} & \multicolumn{1}{c|}{Accuracy} & \multicolumn{1}{c}{Loss} & \multicolumn{1}{c}{Accuracy}\\\hline

\multirow{6}{*}{\rotatebox[origin=c]{90}{FCNN}\hspace{.1cm}}&F(800)	& Sigmoid&	0.2	&0.763	&0.716&	0.845&	0.731\\
&F(400)	&Sigmoid&	0.2&	1.163&	0.758&	0.991&	0.762\\
&F(200)	&Sigmoid&	0.2&	1.076&	0.758&	0.988&	0.761\\
&F(800)/F(800)&	Sigmoid&	0.2&	1.718&	0.759&	1.076&	0.765\\
&F(400)/F(400)	&Sigmoid&	0.2&	1.456&	0.757&	1.075&	0.762\\
&F(200)/F(200)	&Sigmoid&	0.2&1.087&	0.755&	0.928&	0.764\\\hline
\multirow{3}{*}{\rotatebox[origin=c]{90}{LSTM}\hspace{.1cm}}&L(800)&	tanh&	0&		0.772&	0.732&	0.533&	0.739\\
&L(400)	&tanh&	0&		0.697&	0.691&	0.603&	0.706\\
&L(200)	&tanh&	0&		0.555&	0.732&	0.544&	0.741\\\hline
\multirow{4}{*}{\rotatebox[origin=c]{90}{LSTM+A}\hspace{.1cm}}&L(800)/A&	tanh&	0&	1.038&	0.785&	0.602&	0.794\\
&L(400)/A&	tanh&	0&		1.198&	0.758&	0.491&	0.774\\
&L(200)/A&	tanh&	0&		1.004&	0.783&	0.684&	0.786\\
&L (100)/A&	tanh&	0&	0.875&	0.777&	0.805&	0.798\\
\hline\hline
\end{tabular}
\caption{\label{tab:table3}Comparison of intensity classification results for GFCC and MFCC representations.Here F(800) denotes a neural network with one 800-unit fully connected hidden layer. F(800)/F(800) denotes a network with two 800-unit fully connected hidden layers. L(800) denotes a network with an 800-unit LSTM layer. L(800)/A denotes a network with an 800-unit LSTM layer with Attention.}
\end{table}

\subsubsection*{Acknowledgments}
We gratefully acknowledge Peter Lowen for his encouragement, Laurie Heller for her recommendations on the psychology and physiology of speech and hearing, and Google Colaboratory for providing GPU resources.

\printbibliography
%

\end{document}